\documentclass[manuscript]{acmart}

\AtBeginDocument{%
  \providecommand\BibTeX{{%
    \normalfont B\kern-0.5em{\scshape i\kern-0.25em b}\kern-0.8em\TeX}}}

\copyrightyear{2021}
\acmYear{2021}
\setcopyright{rightsretained}
\acmConference[CHI '21]{CHI Conference on Human Factors in Computing Systems}{May 8--13, 2021}{Yokohama, Japan}
\acmBooktitle{CHI Conference on Human Factors in Computing Systems (CHI '21), May 8--13, 2021, Yokohama, Japan}\acmDOI{10.1145/3411764.3445321}
\acmISBN{978-1-4503-8096-6/21/05}

\begin{document}


\title{``I Choose Assistive Devices That Save My Face''}
\subtitle{A Study on Perceptions of Accessibility and Assistive Technology Use Conducted in China}

\renewcommand{\shorttitle}{A Study on Perceptions of Accessibility and Assistive Technology Use Conducted in China}

\author{Franklin Mingzhe Li}
\affiliation{
  \institution{Carnegie Mellon University}
  \city{Pittsburgh}
  \state{Pennsylvania}
  \country{USA}
}
\email{mingzhe2@cs.cmu.edu}

\author{Di Laura Chen}
\affiliation{
  \institution{University of Toronto}
  \city{Toronto}
  \state{Ontario}
  \country{Canada}
}
\email{chendi@dgp.toronto.edu}

\author{Mingming Fan}
\affiliation{ 
  \institution{Rochester Institute of Technology}
  \city{Rochester}
  \state{New York}
  \country{USA}
}
\email{mingming.fan@rit.edu}

\author{Khai N. Truong}
\affiliation{
  \institution{University of Toronto}
  \city{Toronto}
  \state{Ontario}
  \country{Canada}
}
\email{khai@cs.toronto.edu}

\renewcommand{\shortauthors}{Li et al.}

\begin{abstract}
  Despite the potential benefits of assistive technologies (ATs) for people with various disabilities, only around 7\% of Chinese with disabilities have had an opportunity to use ATs. Even for those who have used ATs, the abandonment rate was high. Although China has the world's largest population with disabilities, prior research exploring how ATs are used and perceived, and why ATs are abandoned have been conducted primarily in North America and Europe. In this paper, we present an interview study conducted in China with 26 people with various disabilities to understand their practices, challenges, perceptions, and misperceptions of using ATs. From the study, we learned about factors that influence AT adoption practices (e.g., misuse of accessible infrastructure, issues with replicating existing commercial ATs), challenges using ATs in social interactions (e.g., Chinese stigma), and misperceptions about ATs (e.g., ATs should overcome inaccessible social infrastructures). Informed by the findings, we derive a set of design considerations to bridge the existing gaps in AT design (e.g., manual vs. electronic ATs) and to improve ATs' social acceptability in China.
\end{abstract}




\maketitle

\section{Introduction}

China has the world's largest population with disabilities (83 million) \cite{HandicappedChina}, which is twice as large as that of the US \cite{Chinadisabled2014,Taylor2017statistics}. From many Chinese people's perspective, having a disability is linked to past wrongdoings \cite{campbell2011invisibles}, and they view disability as a problem that needs to be ``fixed'' or pitied \cite{susanne2018disability}, which creates a barrier in social interactions between people with disabilities and the general public \cite{campbell2011invisibles}. Prior work has reported that Chinese people with disabilities are largely invisible from the public in both urban and rural areas \cite{Invisibl32:online}, and have limited education and presence in the workplace. Such a large population with disabilities also severely lacks the care offered by trained professionals. For example, China has only 1/185 as many physiotherapists per person as Europe \cite{HandicappedChina}. Given the large number of people with disabilities and the serious shortage of trained professionals that can offer help to people with disabilities in China, assistive technologies are viewed as an appealing solution to assist people with disabilities. However, only about 7\% of the Chinese with disabilities have had an opportunity to use ATs \cite{theSecondNationalSampleSurveyDisability}.

The practices and challenges surrounding AT and design considerations for improving the social acceptability and adoption of ATs have been the subject for many research studies in the past (e.g., \cite{shinohara2011shadow,profita2016effect,McNaney:2014:EAG:2556288.2557092,boiani2019non,asghar2019influence,asghar2019cultural}). However, these studies were conducted primarily of North America \cite{shinohara2011shadow,profita2016effect,McNaney:2014:EAG:2556288.2557092} and Europe \cite{boiani2019non,asghar2019influence,asghar2019cultural}. Given the large population with disabilities in China, the shortage of trained professional caregivers, the low usage rate of ATs, and different cultural contexts from North America or Europe, it is important to examine the practices and challenges surrounding AT acceptability and adoption within China specifically. In this work, we sought to answer the following two overarching research questions (RQs): 

\begin{itemize}
  \item RQ1: What are the practices and challenges surrounding AT use by people with various disabilities?  
  \item RQ2: What are the design factors that influence the adoption and social acceptability of ATs?
\end{itemize}

To answer the research questions, we conducted a semi-structured interview study with 26 participants with various disabilities: eight with visual impairments, eight with hearing loss, eight with motor impairments, and two with cerebral palsy. From the study, we articulated current problems with AT adoption (e.g., misuse of accessible infrastructure, issues with replicating existing commercial ATs) and challenges of using ATs in social interactions (e.g., Chinese stigma). We then revealed existing misperceptions about ATs (e.g., ATs should overcome inaccessible social infrastructures, ATs with more functionalities are better) and compared our findings with misperceptions about ATs in North America \cite{shinohara2011shadow}. Informed by the findings, we further showed a set of design considerations for improving the social acceptability of ATs, and bridging the existing gaps in AT design (e.g., manual vs. electronic ATs, mainstream technologies vs. ATs). By completing a study, similar to previous research, with participants in China specifically, we contribute an understanding of the practices and challenges surrounding AT use by people with various disabilities and provide associated AT design recommendations under the Chinese context.

\section{BACKGROUND AND RELATED WORK}
\subsection{People with Disabilities in China}
\label{Background of People with Disabilities in China}

Despite China's rapid urbanization, the majority of people with disabilities still reside in rural areas \cite{susanne2018disability}. Although China has the largest population with disabilities, people with disabilities are rarely seen in public spaces \cite{Invisibl32:online}. Chinese with disabilities also have limited education compared to other countries \cite{HandicappedChina}. In 2016, nearly 20\% of Chinese with disabilities were either illiterate or had no schooling \cite{susanne2018disability}. Furthermore, Kim et al. compared special education between China and the United States from national educational statistics \cite{kim2019comparison, chinaEducation,USEducation}. Around 48\% of people with disabilities in China went to special schools instead of studying in regular schools with able-bodied students in 2017 \cite{kim2019comparison,chinaEducation}. In contrast, this number was less than 3\% for people with disabilities in the United States \cite{kim2019comparison,USEducation}. 

In terms of employment, China adopted multiple methods, such as employment by proportion, concentrated employment, and non-profit job allocation to support the employment of people with disabilities \cite{guo2014development}. However, only 28\% of Chinese with disabilities were working in 2017 \cite{susanne2018disability}. Due to the existence of prejudices in the workplace, many people from the general public assume that people with disabilities cannot productively contribute to the economic growth or the society, and should have specialized career paths (e.g., visually impaired individuals trained to be massagers) that often separate them from the general public \cite{susanne2018disability}.

People with disabilities in China, for a long time, were referred to as ``can fei,'' a combination of two characters meaning ``incomplete or deficient'' and ``useless.'' 
Starting from the 1990s, people started using the word ``can ji,'' changing the latter character to one meaning ``disease or sickness.'' 
However, the term ``can ji'' implies that people with disabilities have some kind of incurable ailment that renders them abnormal.
Unfortunately, this term is still widely used today even though the term ``can zhang'' has been suggested (replacing the second character with one meaning ``obstacle or barrier'') \cite{Invisibl32:online}. 
Some Chinese parents still consider having a child with disabilities as linked to wrongdoings in the past \cite{campbell2011invisibles}. In sum, it is not uncommon that many Chinese people still hold a stigmatized view that disability is a problem to be ``fixed'' or pitied. In our work, we explore how the traditional Chinese stigma affects the AT adoption in social interactions.

\subsection{AT Adoption}
\label{AT Adoption}

Given the large number of people with disabilities and the shortage of trained professionals that could offer help to people with disabilities in China, ATs can be an appealing solution to help people with disabilities improve daily functioning, enable a person to successfully live at home and in the community, and enhance independence \cite{scherer1996outcomes}. According to the national sample survey of people with disability in China, only 7\% of the population with disabilities in China has ever used an AT \cite{theSecondNationalSampleSurveyDisability}. Within this 7\% of AT users, over 20\% of them abandoned their owned ATs \cite{theDocumentaryoftheServiceofAssistiveTechnologyinMidwestChina}. 
As a result, it is important to understand how people with disabilities use ATs, why they abandon certain ATs, and the challenges that they encounter when using ATs. To investigate the reasons for AT abandonment, prior research explored the social and personal factors that influence AT adoption and usage \cite{phillips1993predictors,deibel2013convenient,kane2009freedom,kintsch2002framework,parette2004assistive,shinohara2011shadow}. To improve the ATs' adoption rate, it is important to involve AT users in the entire design process \cite{phillips1993predictors,riemer2000factors} or empower them to ``DIY'' their own AT devices \cite{hurst2011empowering}. In terms of the AT design, Riemer-Reiss and Wacker \cite{riemer2000factors} conducted a survey study with 115 individuals with various disabilities and concluded that ATs must meet an important functional need to improve the adoption rate, similar to what Kintsch and DePaula found \cite{kintsch2002framework}. Other factors, including frustration tolerance, minimized stigmatization, and willingness to incorporate ATs into daily routine, could help to reduce technology abandonment \cite{kintsch2002framework}. Deibel \cite{deibel2013convenient} further presented a generalized heuristic model for understanding various factors that influence the adoption and usage of ATs, such as device necessity, task motivation, physical effort, and cognitive effort. Different environments, such as workplaces or social interactions, also affect the choice of ATs \cite{carmien2008design,shinohara2011shadow,wahidin2018challenges}. In our work, we introduce AT adoptions in China through factors that affect AT choices and the unique needs for AT customizations.

Mainstream technologies, such as mobile devices, have been explored for accessibility purposes (e.g., \cite{fan2020eyelid2,Li2020iWink,bigham2010vizwiz,guo2016vizlens,kane2009freedom,kane2008slide,kianpisheh2019face,li2017braillesketch,li2019fmt}). However, challenges and concerns surrounding the adoption of mainstream technologies still exist  \cite{carrington2015but,kane2008slide,kane2009freedom,rodrigues2015getting,profita2016designing}. Kane et al. \cite{kane2009freedom} conducted a qualitative two-method study with 20 participants with visual and motor impairments to examine how people select, adapt, and use mobile devices in their daily lives. The study provided guidelines to design more accessible mobile devices (e.g., increasing configurability and contextual adaptation). Furthermore, the evolving needs of users should also be considered to increase the adoption of ATs \cite{rodrigues2015getting}. 

AT adoption conditions also vary in countries with different levels of income \cite{eide2009assistive}. People who live in low-income countries may have limited access to ATs \cite{eide2009assistive} and lack sufficient knowledge and research on ATs \cite{may1999survey}. For example, Rodrigues et al. \cite{rodrigues2015getting} examined smartphone adoption in Western countries and found that some people with visual impairments continue to use their old feature phones with the availability of other ATs. In contrast, people with visual impairments in Bangalore started to switch to smartphones \cite{pal2017agency} due to the lack of existing ATs or old feature phones. In our research, we discuss the similarities and differences of ATs and mainstream technologies adoption for people with various disabilities between China and other countries.

\subsection{Social Acceptability of ATs}
\label{Social Acceptability of ATs}

Researchers found that lacking considerations of users' opinions is one key factor of AT abandonment, in addition to the poor performance of ATs and the change in user needs \cite{phillips1993predictors}. Users' opinions and preferences of ATs might change based on social contexts \cite{shinohara2009blind}.
For example, depending on social contexts, people with disabilities may feel either self-conscious or self-confident when using ATs \cite{shinohara2016self}. 
To understand how social contexts affect AT use, Shinohara and Wobbrock \cite{shinohara2011shadow} conducted an interview study and found existing misperceptions that pervaded AT use: \textit{ATs could functionally eliminate a disability}, and \textit{people with disabilities would be helpless without their ATs}. 
These findings inspired later research to take social interactions into AT design consideration \cite{profita2016effect}. 
To reduce the misperceptions surrounding ATs, researchers proposed participatory design \cite{lindsay2012empathy}, design for social accessibility \cite{shinohara2016self,shinohara2018incorporating}, and collaborative accessibility \cite{bennett2018interdependence,branham2015collaborative,zeng2015pilot}. Moreover, to reduce unwanted attention surrounding ATs, prior research has also advocated integrating accessibility features into mainstream technologies \cite{shinohara2011shadow,naftali2014accessibility}. 
In sum, social acceptability has been demonstrated to be important for AT design. However, prior research was mostly conducted in Western cultures, and cultural background \cite{ripat2011intersection}, level of education \cite{kaye2008disparities} and access to information services \cite{rubin2001race} may affect AT adoption. China has a different social and cultural context for people with disabilities. 
However, there is a lack of exploration on how social interactions affect the uses of ATs in China. Although some past works have explored AT adoption in developing countries from the design and the user's prospectives \cite{pal2017agency}, these research did little to investigate how the social context affects AT adoption and the existing misperceptions of using ATs in these countries. Thus, in this work, we examine whether and to what extent prior reported causes and solutions of social acceptability of ATs apply to the practices and challenges with AT use in China.

\begin{table*}[ht]
\caption{Participants' demographic information.} 
\centering 

\small
\begin{tabular}{|p{1.2cm}|p{2.5cm}|p{0.4cm}|p{0.8cm}|p{2cm}|p{8cm}|} 

\hline 
Participant & Disability & Age & Gender & Occupation & Assistive Technology \\ [0.5ex] 

\hline 
1 & Born with low vision, lost sight 10 years ago & 34 & M & Massager & white cane, PC screen reader, smartphone screen reader, magnifier, Huawei smartphone, iPhone, slate and stylus, radio, blind poker cards \\
 \hline
2 & Cerebral palsy & 19 & M & Student & standing bed, wheelchair, walking frame, iPhone \\ \hline
3 & Spina bifida & 36 & F & Self-employed & car, iPhone, electric tricycle, sporting wheelchair, crutches, wheelchair trailer \\ \hline

4 & Deaf & 36 & F & Community centre staff & iPhone, Xiaomi smartphone, voice-to-text software, lighting doorbell, artificial cochlea, SIEMENS hearing aid \\
 \hline

5 & Congenitally blind & 44 & M & Massager & e-reader, smartphone screen reader, slate and stylus, white cane, iPhone \\
 \hline

6 & Upper-extremity amputations, no forearm & 61 & M & Retired & mechanical artificial arm, electric artificial arm, motorcycle, PC, smartphone \\
 \hline

7 & Lost sight due to medical accident 15 years ago, totally blind & 35 & F & Software dealer & iPhone, Android phones, Nokia phones, PC screen reader, radio, smart home appliances, smartphone screen reader \\
 \hline

8 & Deaf & 30 & M & Unemployed & hearing aid, artificial cochlea, vibration band, OPPO smartphone \\
 \hline

9 & Motor impairment due to spinal cord injury 13 years ago & 40 & M & Information technology & wheelchair, crutches, iPhone, computer, wheelchair trailer, extended clamp \\
 \hline

10 & Spinal cord injury due to medical accident 7 years ago & 25 & F & Call center telephone operator & wheelchair, wheelchair trailer, Huawei smartphone, extended clamp\\
 \hline

11 & Motor impairment caused by polio & 60 & M & Lottery service & crutches, motorcycle, electric tricycle, sporting wheelchair, iPhone\\
 \hline

12 & Congenitally blind & 42 & M & Massager & white cane, book reader, PC screen reader, smartphone screen reader, slate and stylus, Xiaomi smartphone\\
 \hline

13 & Motor impairment caused by polio & 29 & M & Call center telephone operator & crutches, wheelchair, hand-propelled tricycle, electric tricycle, smartphone \\
 \hline

14 & Deaf & 44 & M & Website operator & voice-to-text software (Shenghuo, Xinsheng, Luyinbao), smart watch, iPhone, hearing aid, iPad, PC, lighting doorbell \\
 \hline

15 & Cerebral palsy & 50 & F & Vegetables sales at grocery store & standing bed, wheelchair, crutches, walking frame, iPad\\
 \hline

16 & Deaf & 44 & F & Unemployed & hearing aid, lighting doorbell, voice-to-text software, Android smartphone \\
 \hline

17 & Congenitally blind & 42 & F & Massage instructor & slate and stylus, talking watch, white cane, book reader (Dushulang), smartphone screen reader, Huawei smartphone, PC screen reader, APP (Didi)\\
 \hline

18 & Congenitally blind & 41 & M & Massager & white cane, smartphone screen reader, navigation APP (Baidu Map), slate and stylus, Braille board, Braille book, e-reader, Xiaomi smartphone\\
 \hline

19 & Low vision & 45 & M & Massager & monocular, magnifier, radio, PC screen reader, white cane, e-reader, iPhone, smartphone screen reader\\
 \hline

20 & Motor impairment caused by polio & 43 & M & Information technology & hand-propelled tricycle, extended clamp, motorcycle, car, crutches, smartphone\\
 \hline

21 & Spina bifida & 31 & F & Video editor & wheelchair, electric tricycle, smartphone, APP\\
 \hline

22 & Deaf & 28 & F & Unemployed & hearing aid, iPhone, iPad, voice-to-text software\\
 \hline

23 & Congenitally blind & 44 & M & Massager & Sunshine screen reader, iPhone, white cane\\
 \hline

24 & Deaf after 2 years old & 30 & M & Teacher & hearing aid, Huawei smartphone, voice-to-text software, APP\\
 \hline

25 & Deaf & 44 & M & Community centre staff & hearing aid, lighting doorbell, vibration alarm clock, smartphone, voice-to-text software\\
 \hline

26 & Hearing loss & 40 & F & Teacher & hearing aid, iPhone, voice-to-text software\\[0.5ex] 
\hline 
\end{tabular}

\label{table:demographic} 
\end{table*}

\section{METHOD}
\label{Method}
We recruited 26 people with various disabilities through local disability communities and conducted semi-structured interviews with them to understand their practices and challenges with using ATs. Interview sessions took place at various local disability communities and lasted approximately 60 - 90 minutes. 
Interviews were audio-recorded, transcribed, and translated for further analysis. The whole recruitment and study procedure was approved by the institutional review board (IRB).

\subsection{Participants}

Previous research has provided an understanding of the perceptions and misperceptions associated with ATs. These findings were uncovered through studies focused mostly on people with visual impairments. However, Phillips and Zhao \cite{phillips1993predictors} found that mobility-related ATs had the highest rate of abandonment.
Thus, to better understand the practices and challenges of AT use for people with various disabilities, we interviewed eight participants with visual impairments, eight participants with hearing impairments, eight participants with motor impairments, and two participants with cerebral palsy. To recruit participants, the researchers reached out to the China Disabled Persons' Federation (CDPF), the largest government authorized organization for Chinese with disabilities, to distribute the study advertisement to people with disabilities. All participants were registered population with disabilities in the CDPF. To understand the practices and existing challenges of using ATs, we recruited participants who had experiences with ATs. Participants were between 19 and 61 years old (mean = 38.7, SD = 9.7). Table \ref{table:demographic} shows detailed information regarding the their age, gender, disability, occupation, and the list of ATs that they have used. Participants were compensated 100 CNY after completing the interview.

\subsection{Procedure}

To understand the perceptions of ATs, we adapted the questionnaire from Shinohara and Wobbrock \cite{shinohara2011shadow} and extended it to learn about the participants' perceptions of mainstream technology (e.g., what types of mainstream technologies do they use? What do they like about them? What mainstream technologies do they want to try the most, but are not currently accessible?), and the differences between mainstream technology and AT (e.g., how does mainstream technology help you compared to AT under different circumstances?). 

We first asked participants about their demographic information, the condition of their disabilities, and their experiences with ATs. We then asked them to compare their previous ATs with the current ones that they are using, if they have used multiple versions of functionally similar ATs. We then asked them to share their experiences and feelings when they use their ATs in social and work contexts. Furthermore, we asked participants to talk about any misperceptions about ATs that they had encountered from the general public, any previous misperceptions that they had themselves, and if they feel self-empowered or self-conscious while using ATs. Additionally, we asked participants what ATs they would like to have in the future, what they think are the most important factors of successful ATs, and their perspectives on how ATs compare with mainstream technologies.

\subsection{Analysis Method}
All interviews were conducted in Mandarin by the first author, who is a native Chinese speaker. We audio-recorded all interviews with participants and transcribed the recordings verbatim. We then translated the transcripts into English for analysis. Two coders independently performed open-coding \cite{corbin1990grounded} on the transcripts. In the open-coding process, Cohen's Kappa equals to 0.85 for the inter-rater reliability between the independent coders initially. Then, the coders met and discussed their codes. When there was a conflict, they explained their rationale for their code to each other and discussed to resolve the conflict. Eventually, they reached a consensus and consolidated the list of codes. 
Afterward, they performed affinity diagramming \cite{hartson2012ux} to group the codes and identify the themes emerging from the groups of codes. Overall, we established 10 themes and 21 codes. The results introduced in the next section are organized based on these themes, and we clustered the themes that overlap.

\section{RESULTS}
In this section, we present the AT adoption practices reported by our participants, the challenges they encountered with using ATs in social interactions, and common misperceptions they faced about disabilities and ATs.
\subsection{AT Adoption}

\subsubsection{Choice of ATs}
\label{choice of ATs}
In the interview, we found that participants' choices of AT are affected by many factors, including self-esteem, limited resources for consultation, advice from people with a similar disability, limited space, social infrastructure, and subsidy coverage. 
First, participants felt that ATs signaled the severity of disabilities and thus tended to choose to use particular ATs that could \textbf{signal to others their self-care abilities with minimal assistance from ATs}. For example, participants preferred crutches over wheelchairs because the former indicates that they could still walk to some extent. This behavior relates to the concept of ``face'' in Chinese culture, which is an individual's contingent self-esteem \cite{hwang2006moral}, and affects their decisions on whether to adopt certain ATs. We found that some participants with motor impairments used their crutches until they were warned by their doctors to stop using them due to the risk of spinal scoliosis caused by their body weight. P20 commented on the reasons for continuing to use crutches until 35 years old before switching to the wheelchair:

\begin{quote}
``...You may ask me why I do not just use an electric wheelchair instead of crutches every day; the key reason is that I want to show other people the ability of my abled body parts---especially the upper body. I do not want other people to think that I am a useless person, I want other people to still think that although I lost the control of my legs, I still have the ability to move with my arms and shoulders. Using my crutches helps me to save face when I walk on the street. This is the reason why I continued using crutches until 35 years of age when my doctor told me that I was too heavy to use the crutches...''
\end{quote}

In China, maintaining ``face'' means that ``shameful'' family affairs cannot be disclosed to outsiders. Due to some negative perceptions on disabilities, the family of a child with disabilities may be reluctant to seek supportive services \cite{ravindran2012cultural}. From our study, we found that most participants sought information about what skills they could learn and what ATs they could use by \textbf{primarily consulting family members and friends around them}. However, their family and friends often do not have the same disabilities and cannot offer accurate information about what people with certain disabilities could learn or what ATs would be the best fit for them, which limits their understandings of what ATs they could or should use. Consequently, 88\% of participants reported that they did not know what they could learn or do when they were young, and their primary information source was limited to their family members and friends. In our study, we found that this circumstance in China delayed the process of learning what people who acquired a certain disability later in their life could use to become more independent. We also discovered that some people would start with a very pessimistic view on how to live independently once they acquired a certain disability. \textit{"It took me years to know that I could still work in IT and live independently. There lack professional consultants that I could ask when I just had the accident, and it really made me feel desperate initially"}, said P9. Therefore, it would be beneficial to have consultants who could provide precise and customized advice to people who just acquired a disability regarding what ATs they could use and career planning.

\textit{``I did not know that I could use smartphones until I met a person with a similar impairment who used the smartphone every day''}, continued P9. When people saw that others with the same disability could do certain things, they started to realize that they can use certain ATs too. Participants mentioned that seeing and \textbf{knowing what people with a similar disability can do} gives them the confidence to also learn to use particular ATs. P20 commented that knowing someone with an even more severe motor impairment who can drive a car inspired him to learn how to drive:

\begin{quote}
    ``...When I did not know what I can do, everything was hard for me. However, I realized later that there are so many things I could do. This made lots of changes; I started to become more optimistic about everything. I wish I knew what I could do at the beginning. After I found that other people with even more severe motor impairments could drive cars, I bought mine, and I own two cars now...''
\end{quote}

The \textbf{surrounding space} that participants interacted with also impacted their decisions on what ATs to use. We found that most of the home environments our participants lived in required special accessible modifications to be fully accessible. One potential reason is that many Chinese people live in apartments or condos due to the high density of the population, which is different from people who live in many areas in North America \cite{1Whatper6:online}. Seven participants also mentioned that they have to live in older apartments for the lower rent, but these apartments are not accessible or required further modifications. For some participants, their bathroom size was too small to roll a wheelchair in. P11 mentioned the problem of his bathroom and kitchen:

\begin{quote}
``...Most of the apartments were not initially aimed to be designed for people with disabilities. I found that nearly all of them need some modification, such as increasing the door size or reducing the height of the kitchen stove. Even so, I still cannot move around with my wheelchair in the bathroom. This forced me to use the crutches while I was using the bathroom. But it is more dangerous...''
\end{quote}

In addition to the home environment, participants also revealed that \textbf{inaccessible public places} usually forced people with disabilities to use specific ATs. Ten participants mentioned that they are used to visiting specific accessible public places in the city, even though it may require a long commute. However, they often have to visit unfamiliar places for various reasons, where the accessibility of the new environment is unknown to them beforehand---such as traveling in a different city, hanging out with friends, or visiting a customer. \textit{``I always bring my crutches with me when I visit unfamiliar places because they either do not have an elevator or the washroom is not accessible''}, said P11. Even if participants tried to always visit accessible places, there are still constraints which forced them to stay at inaccessible places. For example, P9 complained about the washroom he experienced at work:

\begin{quote}
    ``...The washroom in my office building is not accessible to me. The door is not even as wide as my wheelchair. More importantly, there is a small step in each unit of the washroom. It forced me to use crutches instead of my wheelchair. Using the washroom at work is the hardest task for me every day...''
\end{quote}

\begin{figure}
\centering
  \includegraphics[width=1\columnwidth]{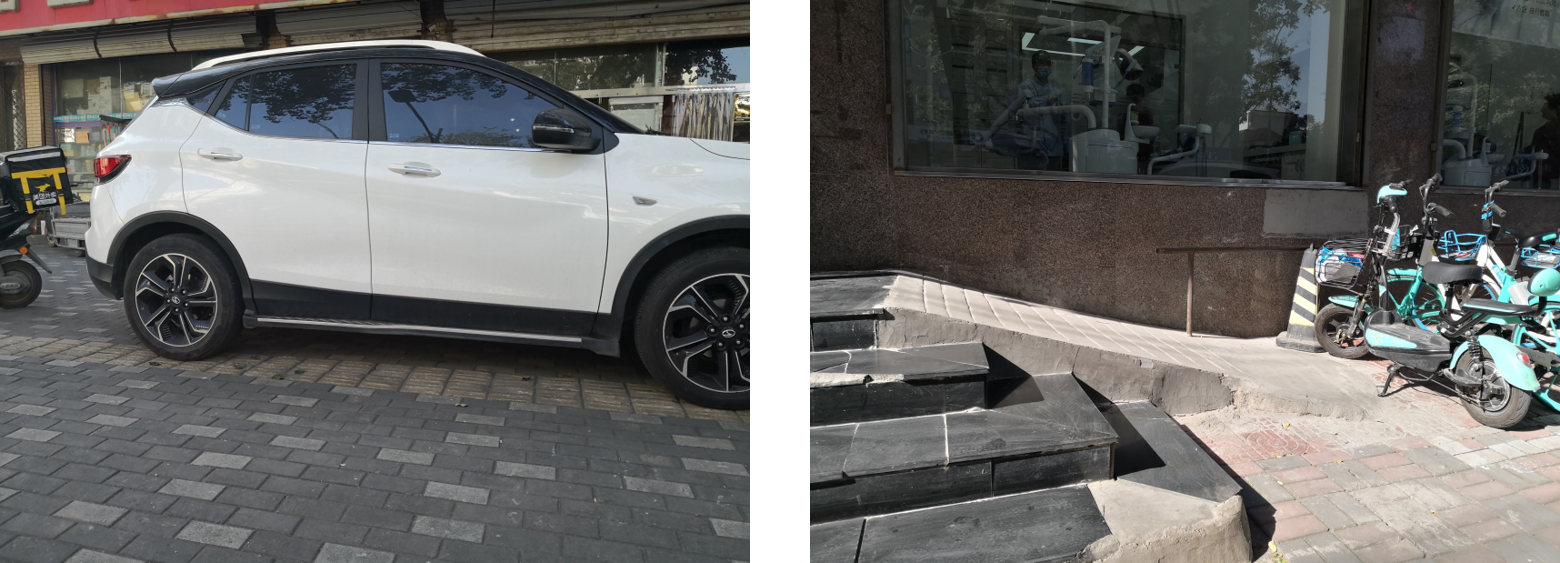}
  \caption{Misuse of accessibility infrastructures by members of the general public. Left: the tactile paving was blocked by the car. Right: the ramp was blocked by locked electrical bikes.}~\label{fig:tactile}
  \Description{There are two pictures. The left picture shows a white car that parked on the tactile paving. The right picture shows a ramp and the lower side of the ramp is blocked by several blue electrical bikes.}
\end{figure}

In our study, we also found that social infrastructures that are supposed to help people with disabilities engage in social interactions, such as the tactile paving and the wheelchair ramp, are widely constructed in public. However, participants reported that the \textbf{misuse of accessibility infrastructures by members of the general public} posed even more safety concerns than without any infrastructures, which forced people to use certain ATs. For example, people locked their bikes on the ramp or placed random objects on the tactile paving (Fig. \ref{fig:tactile}). 

Finally, financial conditions varied among people with disabilities and affected their choice of ATs. Most participants mentioned that they could not get their desired ATs because of financial concerns. 
Other than general financial conditions, some participants who have insurance or subsidy to buy ATs explicitly mentioned that the \textbf{subsidy coverage is often restricted to certain brands or models of ATs}, which may not necessarily match their needs. P8 commented on this:

\begin{quote}
``...My current hearing aid is really ugly and always have a loud noise that annoys me all the time, I know there is a [brand of] hearing aid that is much lighter and better designed than my current one, but it is too expensive, and my insurance does not cover the cost for that brand...''
\end{quote}

Beyond the subsidy to buy ATs, we found that the \textbf{high cost of the maintenance fee is not covered by the subsidy}, which also affects the choice of ATs. Similar to what Armstrong et al. \cite{armstrong2007evaluation} found in Afghanistan, even if people obtained certain ATs through subsidy or donation, the AT maintenance problem and the lack of replacement parts still affect the adoption of ATs in developing countries. \textit{``I can use my subsidy to buy my electronic artificial arms, but it is too expensive to repair it, and my subsidy does not cover the maintenance cost. That is one of the reasons for using my mechanical artificial arms now''}, said P6.

\subsubsection{``DIY'' and Customized ATs}
\label{``DIY'' and Customized ATs}

We found that most participants used ATs that were constructed by themselves, family, or friends (Fig. \ref{fig:figure1}). From the interview, we observed two main practices of ``do-it-yourself'' (DIY) ATs: replicating existing commercial ATs and modifying inaccessible mainstream technologies. Unlike the purpose of ``DIY'' ATs in North America \cite{buehler2015sharing,hurst2011empowering}, which leveraged fabrication tools to create new customized ATs or to make AT functional attachments, many of the participants' customized ATs aimed to \textbf{replicate existing commercial ATs} (Fig. \ref{fig:figure1}(a,b)). Different from using 3D printers or other fabrication tools in North America, most of the customized ATs that we learned about were created through handcrafted or traditional ways. For example, crutches or canes were crafted from wooden sticks (Fig. \ref{fig:figure1}(a)). The key reasons behind it are financial concerns, lack of AT designers, and the knowledge gap about the available ATs and how to use them. P13 commented on a hand-propelled tricycle made by his father:

\begin{figure*}
\centering
  \includegraphics[width=1.0\linewidth]{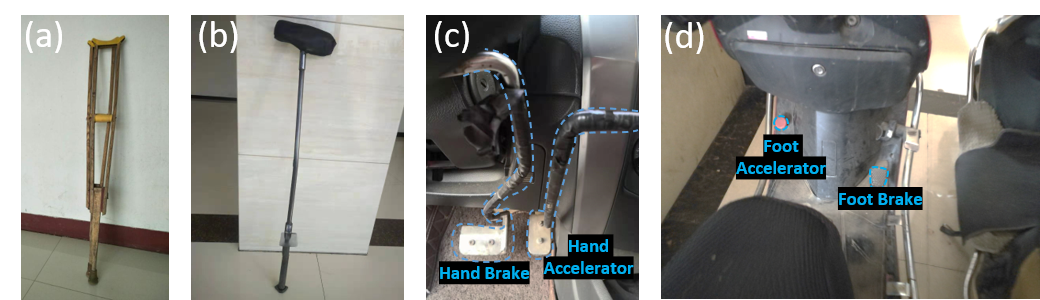}
  \caption{
  (a,b) Self-made crutches which lack careful design and engineering considerations. (c) Accessibility modifications made to a car by a third-party company for a participant with motor impairment. (d) Self-modified brake and accelerator of the motorcycle of P6 (no forearm).}~\label{fig:figure1}
  \Description{There are four pictures in a row. The first picture shows a pair of self-made crutches that is made of wood. The second picture shows a self-made crutch. The third picture shows a modified car hand brake and hand acceleration. This picture has blue dash lines and words with blue font to indicate the shape of modified hand brake and hand acceleration. The fourth picture shows a foot acceleration and a foot brake on a motorcycle. The shape and the location of the foot acceleration and the foot brake are marked and labeled in blue font.}
\end{figure*}

\begin{quote}
``...When I was young, I used to craft my `crutches' from the wood found in the forest in our rural village. One of the key reasons was the high cost of a wheelchair or crutches. Later, my dad modified and made a hand-propelled tricycle for me due to the slow movement of my `crutches.' Other than financial considerations, none of my friends or family have the same impairment as me, which made it hard for me to know what assistive devices suited me better...''
\end{quote}

Although ``DIY'' ATs were functional to some extent and were often economical as reported by participants, these ``DIY'' ATs typically lacked careful design and engineering considerations, and therefore often posed health risks. For example, using crutches with an inappropriate length over time can cause lumbar spine distortion or periarthritis. P11 commented on his ``DIY'' crutches:

\begin{quote}
    ``...When I was young, there did not exist anything called assistive technologies; all I used was just wooden crutches crafted by my parents. The left crutch was slightly shorter than the right one. It made my left shoulder feel really painful when I used it for a long time. Later, my doctor told me that I had periarthritis. Even now, it has never recovered...''
\end{quote}

Beyond constructing ATs from scratch, people also \textbf{modified some inaccessible mainstream technologies} (Fig. \ref{fig:figure1}(c,d))---such as cars and motorcycles---to make them accessible. 
Participants reported that most of the automobile companies do not support any modifications for accessibility in China. As a result, they had to ask their friends or other third-party companies to modify their vehicles. Such modifications, however, are often done by non-professionals and pose safety concerns for both users and the general public. Fig. \ref{fig:figure1}(c) shows the modification of a participant's car by adding hand controllers for the brake and the accelerator. 
Similarly, P6 asked one of his friends to add a foot-controlled brake and accelerator onto his motorcycle to overcome the loss of his forearms. Although these modifications functionally allowed people to use these devices, the poor quality of the modifications and the lack of engineering considerations could potentially be dangerous for people who use them. P6 commented on his experiences and safety concerns of his motorcycle:

\begin{quote}
    ``...I like my motorcycle; it allowed me to visit different places. However, it took me lots of effort to get it modified. Initially, I visited the original motorcycle company, and they told me that they did not offer any type of accessible modifications to their products. Then, I talked to one of my friends, who was a mechanical technician. He then modified my motorcycle and added the red button to allow me to accelerate with my foot. I think the design and the placement of the brake and accelerator need to be improved. I got injured in the past when I tried to reach the foot accelerator, which might be too high for me, and it caused the whole motorcycle to become unbalanced. I am glad the speed was not too fast, but it still made me fall off the motorcycle and bruise my leg...''
\end{quote}

\subsection{Challenges with Using ATs in Social Interactions}

\subsubsection{Stigmatization}
\label{Stigmatization in China}

Participants discussed the negative impact of stigmatization of using ATs, which could have been caused by Chinese traditions, infrastructures, and the knowledge gap from the general public. In traditional Chinese culture, the Buddhist belief of karma caused the negative perspectives on disabilities: it is regarded as punishment for the parents of people with disabilities or past life sins \cite{ChineseC80:online}. The negative perspectives give social stigmas to people with disabilities through social interactions. Currently, \textbf{traditional Chinese stigmatizing terms} are still being used to refer to people with disabilities. For example, ``long zi'' refers to people who are deaf or hard of hearing, ``xia zi'' refers to people with visual impairments, and ``que zi'' refers to people with motor impairments. These traditional terms were used throughout Chinese history and have derogatory connotations and sound humiliating to people with disabilities. 96\% of participants recalled having past memories of being referred to using these terms. P7 described her feelings when she heard a conversation between a mom and a son beside her about her blindness:

\begin{quote}
    ``...I was walking on the street, and I did not say anything or ask anyone for help. Maybe I was walking a little bit slow. There was a mom and a son beside me; the son asked his mom about why I walked slow. I heard the whisper from the mom: `she is a `xia zi,' she cannot see us.' I was really depressed because `xia zi' sounds like I did something wrong and I am a useless person...''
\end{quote}

Participants mentioned the high usage of these derogatory terms in some slang and public shows, which posed difficulties to eliminate the use of these terms in the general public. Beyond being called derogatory terms, participants recalled being stigmatized by others who call attention to and limit their AT usages. P13 described his embarrassment of being called out by the subway station general announcement to turn off some functions of his wheelchair trailers throughout the whole subway station:

\begin{quote}
    ``...I remember that I got called out in the subway, and it made me feel really embarrassed. A staff announced on the public speaker: `the person with the electric wheelchair, please turn off your automatic functionalities when you are on the train.' This made all other people stare at me, and I felt really self-conscious...''
\end{quote}

It can be perceived as good intent by the general public to allocate \textbf{designated areas} for people with disabilities. However, such settings may make people with disabilities more self-conscious. For example, in movie theatres, people with motor impairments are limited to sit or park their wheelchairs in a front area where everyone else can see them. Participants typically wanted to blend into the general public when they are in public settings and did not want to be called out or draw people's attention. P20 commented on this: 

\begin{quote}
    ``...I like that some places now have accessible areas for people with disabilities. However, it is still restricted to a certain area. These areas are either at the front or at the entrance. I found that it caused a lot of attention...''
\end{quote}

The existing stigmatization from the general public affected people with disabilities in social interactions and also influenced their choices of using their AT devices.

\subsubsection{Employment challenges with using ATs}
\label{Employment challenges with using ATs}

In general, we found that our participants encountered various challenges from employment, such as transportation to work, the inaccessible working environment, and unwanted attention. In our study, 65\% of our participants chose to work where the majority of employees have similar disabilities. For example, P5 and P18 worked at a massage clinic called ``mang ren an mo'' in Chinese, which means ``blind massage.'' All the employees in that clinic are people with visual impairments. Another example is a cafe where P16 used to work at, where all employees are people with hearing loss. This \textbf{clustering effect} reduced some of the concerns of unwanted attention or the inaccessible working environment for people with disabilities. However, it may also reduce the interactions between people with disabilities and the general public, which may further cause misperceptions.

In China, the State Taxation Administration \cite{EmploymentInChina60:online} has the policy of tax reduction if a company employs over a certain number of people with disabilities. However, our participants mentioned that they still have a hard time finding a job where the majority of employees do not have the same disability. We found that participants with various disabilities had some difficulties using their ATs while working due to two potential problems: \textbf{feeling self-conscious due to their unique office workstation setup}, and \textbf{incompatibility of accessibility features on work devices}. P9 talked about his concerns about the incompatibility between his wheelchair and the office desk:

\begin{quote}
``...My ATs do not affect me much during work. As far as I know, for most people with motor impairments, like me, our work mostly relies on our upper body. The only part that made me uncomfortable is the height of my table at work. As you know, most of the wheelchairs are not capable of adjusting the height. And we are shorter than other people who sit on a normal chair, that made me really uncomfortable, and my manager bought me a new desk with a lower height. However, it made me look different in the company; I felt self-conscious when other people walked by my desk...''
\end{quote}

P7 reported the problems of using different screen readers that made her customers lose patience:

\begin{quote}
``...My work required me to use my computer to record customer and product information. At the first time, I was trying to use my screen reader on the company's computer. Since my screen reader only supported an old Windows system and was not compatible with the system that my company used, I ended up using another screen reader from another software company. Different computer systems and screen readers really delayed my work, and my customer complained to my manager about that...''
\end{quote}

\subsubsection{Knowledge Gap on ATs}
\label{Lack of Knowledge}
In this section, we further elaborate on the existing knowledge gap between people with disabilities and the general public, and the associated consequences (e.g., misperception, unwanted help). In China, around 48\% of people with disabilities went to \textbf{specialized schools} instead of regular schools; however, this number is less than 3\% in the US \cite{kim2019comparison,chinaEducation,USEducation}. Most of our participants with congenital disabilities went to specialized schools for professional skills training, such as massage. This separation of schooling caused mutual misperceptions between people with disabilities and the general public. Participants mentioned that the general public's lack of knowledge on accessibility sometimes posed threats and dangers to people with disabilities. For example, the misuse of tactile paving is dangerous to people with visual impairments which prevent them from walking on the street.

Due to the separation of education between people with disabilities and the general public, the general public lacks understanding on how to offer help appropriately. Therefore, some people tried to offer \textbf{unwanted help} which may lead to safety concerns. For example, some people pushed the wheelchairs from behind without being asked to do so, which can be very dangerous. P20 mentioned his experiences when he was on the street with his wheelchair:

\begin{quote}
    ``...first, I want to say that I am delighted that other people offered me help. However, they lacked some basic knowledge. For example, many wheelchairs do not require someone to push from the back. If you push someone with a wheelchair, they may feel uncomfortable, and it might be dangerous when there is a bump...''
\end{quote}

In addition, some people may choose to lift a wheelchair to a bus without asking for their user's permission. This could have a potential negative impact on the wheelchair user's dignity. ``I know other people wish to help me, but being lifted in public made me feel so bad and embarrassed,'' said P20. It suggests that although many ATs are viewed primarily as personal devices, some ATs also have a ``social'' aspect that invites people to offer help. For example, wheelchairs are often used by people with disabilities and their caregivers in many social settings, such as hospitals and airport terminals. Thus, when designing ATs, the social aspect of ATs should be considered so that their designs offer clear affordance and signals to invite proper use.

\subsection{Misperceptions about Disability and ATs}
\label{Misperception}

Previous research has identified several misperceptions under a North American context: ATs could functionally eliminate disabilities, and people with disabilities would be helpless without their ATs \cite{shinohara2011shadow}. In this section, we present the similarity of the misperceptions and new findings from this study (e.g., ``ATs should overcome inaccessible social infrastructures'' and ``ATs are symbols of permanent disabilities'').

\subsubsection{``ATs could functionally eliminate the disability'' and ``people with disabilities would be helpless without their ATs''}

Similar to Shinohara and Wobbrock's findings \cite{shinohara2011shadow}, the misperception ``ATs could functionally eliminate the disability'' also exists in China for ATs used by people with motor impairments, hearing impairments, visual impairments, and cerebral palsy. For example, participants pointed out that other people thought wheelchairs could allow people with motor impairments to move freely. However, even with the wheelchair, there are still many obstacles, such as having a hard time entering narrow spaces and climbing stairs. Notably, many participants with hearing impairment mentioned misperceptions related to the hearing aid---that it allows them to hear the sound clearly. However, they claimed that most of the hearing aids could only allow them to recognize whether there is a sound or not. P14 felt annoyed when other people tried to speak close to his ear very loudly and repeat multiple times:

\begin{quote}
    ``...Having a hearing aid does not mean that I can hear the conversations of other people. I can use the hearing aid to locate the sound source, but not understand what a person is talking about. I felt extremely annoyed when some people tried to repeat something to me multiple times and with increasing volume! All I can hear is the noise! I just could not understand, and it annoyed me...''
\end{quote}

Furthermore, participants mentioned that some people have the misperception that cutting-edge healthcare systems should completely eliminate disabilities. According to P2, \textit{``When I was having my lunch at the school cafe, I heard some students whispering: `our healthcare system is way more advanced than 30 years ago, why are there still people with disabilities?'"}. P2 continued: \textit{``because people with disabilities are largely not visible in public due to inaccessible infrastructures, the general public have fewer opportunities to learn about and understand our lives.''} This situation made the misperception hard to resolve.

Beyond the misperception that ``ATs could functionally eliminate the disability,'' our findings agree with another misperception that ``people with disabilities would be helpless without their ATs'' \cite{shinohara2011shadow}, which was found in the Western context. Our participants with visual impairments mentioned their experiences of attracting unwanted attention when using their mainstream devices. \textit{``They were surprised that I could use iPhone X to call an Uber! Some people said I could not do anything without my cane. However, I can use my iPhone to do lots of things without me physically being there''}, said P7. Moreover, we found that people with disabilities tend to rely on specialized ATs less often than before due to the introduction of mainstream technologies with accessibility features. Our participants are already using their smart devices to order food delivery, shop online, contact friends and enjoy entertainment without the need for ATs, such as canes or wheelchairs.

\subsubsection{ATs should overcome inaccessible social infrastructures}

\begin{figure}
\centering
  \includegraphics[width=1\columnwidth]{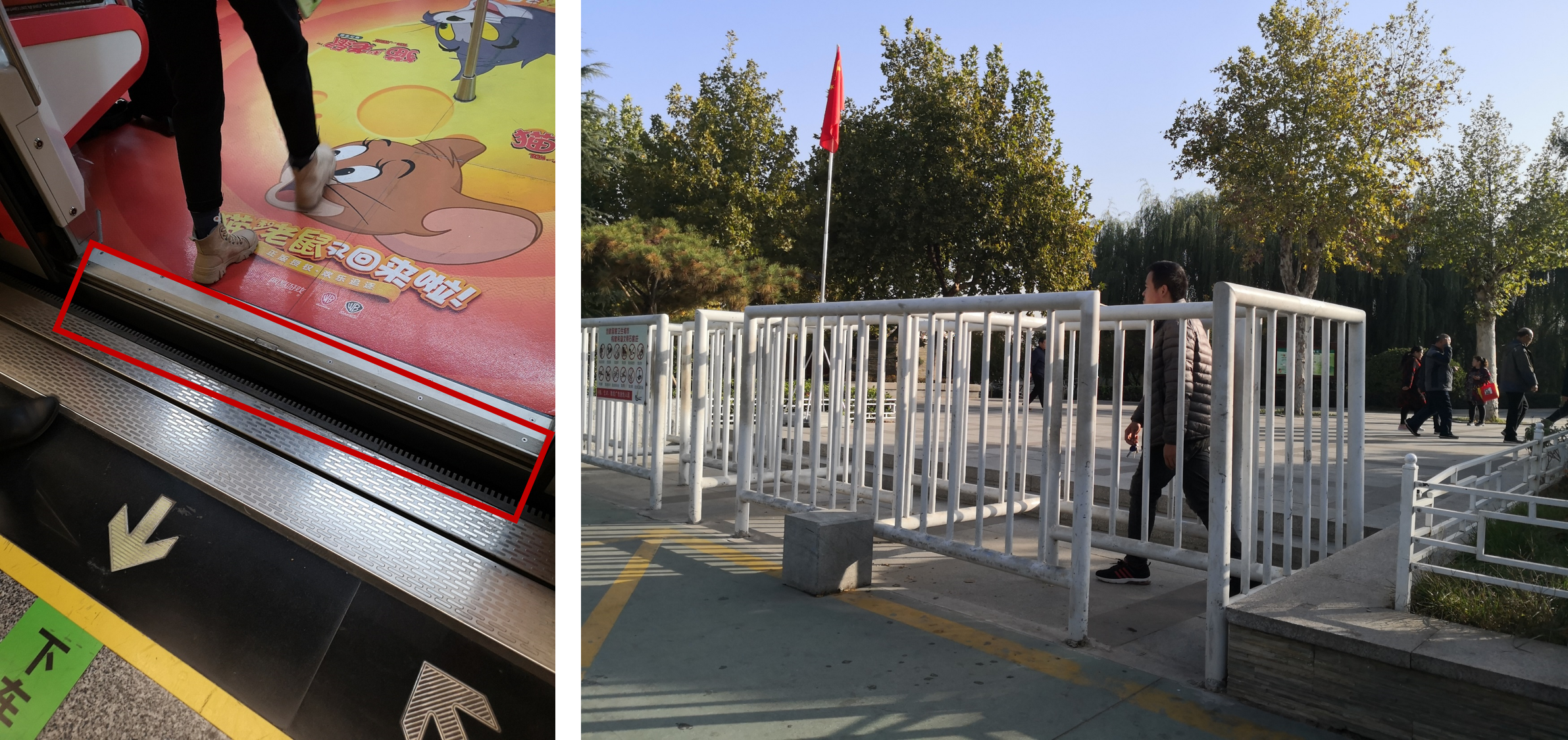}
  \caption{(Left) Local subway system has a gap of over 100 mm between the train and the platform which can decrease the accessibility of the subway system to wheelchair users; (Right) the entrance gate of a park blocks wheelchair users. }~\label{fig:Subway}
  \Description{There are two pictures. The left one shows a subway and the platform. There is a red rectangle that marks the gap between the train and the platform. The right figure shows a white entrance gate to a place with lots of trees. In the picture, one person is walking in the gate.}
  
\end{figure}

Making infrastructures universally designed or modified \cite{iwarsson2003accessibility} for accessibility purposes would help eliminate the need for designing ATs, which are often secondary solutions to inaccessible infrastructure. 
For example, public transit systems in North America, such as the Toronto subway, were built with accessibility issues considered, which requires that the horizontal gap between the subway train and the platform be less than 89 mm for accessibility purposes \cite{GapBetwe50:online}. 
In contrast, some subways in China were not built with accessibility in mind, having gaps of over 100 mm between the train and the platform (Fig. \ref{fig:Subway} Left). As P11 expressed, \textit{``I hate the local subway; my wheelchair just cannot move through the gap. When I talk to people, most of them focused on how I should modify my wheelchair rather than how to make the subway system fully accessible.''}

Furthermore, some participants commented that the general public in China thinks modifying social infrastructures will take more effort than modifying individual ATs to adapt to social infrastructures. In particular, they mentioned that many of the old buildings were not designed to be accessible when it was built, and the building structure may need to be changed to make it accessible (Section \ref{choice of ATs}). \textit{``Some people think that individuals with disabilities should compromise or even sacrifice for the society''}, said P9. Overcoming different social structure barriers by modifying ATs might eventually make the ATs more complicated.

The design of certain park entrances is another example of the challenge involved in overcoming inaccessible social infrastructures. For instance, the S-shaped metal gates of the park entrance (Fig. \ref{fig:Subway} Right) was intended to block vehicles or bicycles from entering, but also accidentally blocked out people with disabilities who use wheelchairs. 

\subsubsection{ATs are symbols of permanent disabilities}

We found that the general public tends to assume that people who use ATs have permanent disabilities. This is a misperception because many people who use wheelchairs may recover from their temporary disabilities. People with broken legs or any other injuries with the lower body may require a wheelchair. Furthermore, participants reported that people thought that if they use certain ATs, they would need it for the rest of their lives. P15 commented on this:
\begin{quote}
    ``...I use my standing bed and walking frame every day to recover from my disability. Some people thought I might need them for the rest of my life. However, once my condition gets better from rehabilitation, I may switch to new ATs that give less assistance...''
\end{quote}

This misperception points to the importance of designing for people's practical needs. \textit{``People with temporary impairments do not need to think about independence, but we do''}, commented P20.

\subsubsection{ATs with more functionalities are better}

Participants reported that the general public often thinks that ATs with more functionalities are better for people with disabilities. For example, participants with motor impairments mentioned that they were asked why they did not just use a multi-functional electric wheelchair. As we introduced before, people might prefer using certain ATs over others because of the following factors: signaling to others their remaining abilities and financial considerations (Section \ref{choice of ATs}). P20 mentioned the importance of considering practical needs in the ATs design:

\begin{quote}
    ``...I found that so many existing ATs try to add as many fancy functions as possible. Do we really need them? These manufacturers really need to think from the user's perspective...''
\end{quote}

\section{Discussion}

In the results section, we described 1) the unique findings of AT adoptions practices employed by people with various disabilities in China (e.g., factors that affect the choice of ATs, ``DIY'' ATs), 2) challenges with using ATs in social interactions (e.g., stigmatization and employment challenges), and 3) existing misperceptions about disabilities and ATs in China (e.g., ATs should overcome inaccessible social infrastructures). Based on our findings, we present the following questions for researchers and designers to consider pertaining to challenges surrounding AT adoption and design. 

\subsection{How Can Manual and Electronic ATs Be Designed Differently to Help People with Disabilities?}

From the interview, we found that there are roles for both manual and electronic ATs, and people with disabilities need both of them in their lives (Section \ref{choice of ATs}). Some participants mentioned the existing misperception from the general public in China that it is more preferable to use more electronic ATs and less manual ATs. We found that both manual and electronic ATs have their own situational uses. 
Electronic ATs may have the benefits of faster movement speed and improved safety. However, people with various disabilities chose to also use manual ATs. Potential reasons include financial concerns, exercise, and signaling to others their remaining abilities to maintain ``face'' as discussed in section \ref{choice of ATs}. 

When designing electronic ATs, it is essential to ensure that the appearance of ATs can communicate to others the abilities of their users.  For example, the wheelchair could be designed with foldable components to support users to stand up briefly if they could do so, and be able to signal this affordance to others clearly. It is also important to design electronic ATs to encourage people with disabilities to exercise as much as possible. For example, potential electric wheelchairs could be designed to force the users to roll the wheelchair for more than a certain number of strokes before they can turn on the automatic function.

In our interview, we found that participants with motor impairments preferred using manual wheelchairs over electric wheelchairs. However, a recent article revealed that the general public might view a manual wheelchair as a more stigmatizing AT than an electric wheelchair in Norway \cite{boiani2019non}. Boiania et al. \cite{boiani2019non} found that the general public has more negative perceptions towards manual wheelchairs in terms of comfortability, aesthetics, and enjoyability. Therefore, designers should take feedback from both people with motor impairments and the general public into consideration, such as using participatory design or co-design methods. 

In terms of manual ATs, we found that each device is mainly used for a single situation or purpose. For example, the cane is always and only required when people with visual impairments needed to walk outside. Unlike manual ATs, electronic ATs are more centralized---smart devices become multi-functional. People with visual impairments used to carry a book reader and a radio to acquire information, and a slate and stylus to share their own thoughts with other people. As smartphones became more accessible, people with visual impairments can simply install screen readers on their smartphones to accomplish the tasks that they may have needed several devices to accomplish before.

In our study, participants with motor impairments mentioned the misperception that ``more functions are always better'' in ATs design. For example, P9 mentioned the experiences of unwanted pushing from the back, and he removed the handle at the back of the wheelchair. This problem does not only exist in China; Low \cite{Spikesan38:online} reported that a wheelchair user from the UK put spikes on the handles to prevent the unwanted pushing. Furthermore, some participants commented on the bulky design of some multi-functional ATs. We can conclude that multi-functional ATs are not always better than the ones with a single function. ATs designers should take into consideration whether or not add-on functions will cause unwanted interactions. On the contrary, we found that most participants with visual impairment preferred having more functions on their canes, such as integrating the navigation function on the cane. We see that people with different disabilities may have varying preferences on functionalities of different ATs.

\subsection{How Should Customization Be Integrated into AT Design under the Chinese Context?}

In China, people with disabilities generally lack support from trained professionals \cite{HandicappedChina}; it is hard for them to find a professional AT designer for ATs that suit their special needs. As 3D printing and quick prototyping technologies are being increasingly used in many aspects of our daily lives, these techniques might allow people with disabilities to design their ATs by themselves. As the costs for personal fabrication technologies keep decreasing, it might be possible to consider how to enable people with disabilities to leverage technologies to customize their ATs \cite{reichinger2018pictures,buehler2014abc,kane2014tracking,baldwin2017tangible}, especially how to enable these fabrication technologies in rural areas where AT users lack access to technologies. Furthermore, collaborative design of ATs by empowering people with disabilities to design for themselves may reduce concerns of functionalities and aesthetics \cite{branham2015collaborative}.

Participants commonly mentioned that their ATs lacked design considerations of where and when these devices are used. This includes using ATs indoors, outdoors, at daytime, at night, and by people with various conditions of disability. To incorporate ATs into daily routines to improve AT adoption \cite{kintsch2002framework}, the AT designers should take where the user may use the ATs into consideration. For example, users might use crutches outside when the ground is slippery with ice. Beyond different locations, users may also use the same AT during the daytime and at night. P18 mentioned that he was hit by bicycles several times when he used his cane at night. Although some of the canes have reflective strips, it may not be enough for people to use canes in dark environments.

Therefore, it is important to take the user's contextual information into consideration while designing ATs. In future work, it would be interesting to create a set of design metrics on how different contextual information may affect design decisions. Once we collect the users' individual AT usage contexts, we could customize their ATs based on the usages.

\subsection{How to Help People with Disabilities Improve Their Understandings of Disabilities, Available ATs, and Career Opportunities?}
We observed the practice of consulting family members and friends on what ATs to use and the benefits of knowing what other people with a similar disability are using (Section \ref{choice of ATs}). There are now over 83 million Chinese people with disabilities \cite{HandicappedChina}, but online resources with related information are lacking. Current online information sources tend to provide general information, such as what a person with spinal cord injuries should and could do. However, different people may have different severity of disabilities. In the interview, 19 participants mentioned that seeing what their peers use is the most common approach to learn about what ATs are available and how to use them.
These participants all commented that it is tremendously helpful for them to learn what they can do and how they can use certain ATs from a larger community where people with similar disabilities are setting examples. In China, there are very few online platforms for the population with disabilities to acquire information related to their disabilities. More importantly, most of these platforms \cite{ChineseDPWeb33:online} are operated by the government and are often designed to increase the public's knowledge about the disabilities rather than supporting interactions among users. Interactive platforms, such as StackOverflow and Github, allow people with common interests but different levels of expertise to interact and share their experiences and questions, which benefit the growth of users. 
On the other hand, we also see such communities for people with visual impairments emerging, such as on Reddit \cite{Accessib39:online}. Creating channels for people with disabilities to know what their peers do could potentially open new opportunities for them and make them feel self-empowered and encouraged by the excellent performance of their peers.
At the same time, due to widespread stigma, it is also challenging to design such online communities so that people with different levels of disabilities would feel comfortable to share their experiences and questions without being judged.  

In addition to government support, the disability movement and Disabled Persons' self-help Organizations (DPOs) have recently begun to emerge in China \cite{zhang2017disabilitysociety}. In addition to supporting DPOs in promoting social and policy changes, our findings also show that it is worth considering to build online platforms that allow people with disabilities to share their successful stories and what they can do with their peers so that the community can inspire each other and allow them to understand potential opportunities that they would otherwise have no access to. More research should be conducted to understand the features, functions, and the interaction mechanisms that such platforms should provide.

Emerging types of social media may also provide more opportunities and resources to share information on ATs and career development among people with various disabilities. Recently, live-streaming platforms have been explored to share knowledge \cite{lu2018you} and even promote intangible cultural heritage \cite{lu2019feel}. Live-streaming has fewer physical constraints, which may cause fewer mobility concerns for people with disabilities, and the monetary gifts that the live-streaming audience sends may generate additional income to cover their daily costs. In the future, it is worth exploring ways to leverage existing live-streaming social media platforms or create new live-streaming social media platforms for people with disabilities to share their stories and to potentially reduce misperceptions held by people with disabilities and the general public.

\subsection{How Can Mainstream and Emerging Technologies Be Leveraged to Improve AT User Experience?}

As smart IoT devices play a more prominent role for people with disabilities in China, participants started to reduce or even replace their traditional ATs. For example, participants with visual impairments commented that they used the cane less often after they could order food delivery and shop online on their smartphones. Our findings verified Shinohara and Wobbrock's \cite{shinohara2011shadow} predictions around 2011 on the trend that people with disabilities would use more mainstream technologies with accessible features. In Shinohara and Wobbrock's \cite{shinohara2011shadow} study, more than half of the participants did not use smartphones. However, we found that all of our participants now use different kinds of smartphones for daily purposes. 

Participants with different disabilities mentioned the frequent uses of various smart devices, such as smart speakers, smartphones, smart curtains, and smart rice cookers. We found that participants used smart devices to mostly complete Instrumental Activities of Daily Living (IADL) \cite{Learning73:online}, such as shopping, paying bills, and cooking, by reducing the effort of movement, searching, and physical actions. Most participants expressed strong interest in trying new smart IoT technologies. This generates opportunities and concerns for smart IoT designers to consider when designing accessible smart IoTs. We found that participants have an open mind about new technologies and innovations. For example, participants with visual and motor impairments showed strong interest in self-driving vehicles.
The enjoyment of smart technologies and the open-mindedness for emerging technologies by people with disabilities in this study is different from the findings of the recent research conducted in Bangalore (India) that showed people with disabilities there adopted smartphones over feature phones because they had no choice due to the increasing market share of smartphones in the society \cite{pal2017agency}. It would be interesting to examine what caused this different attitude toward smart technologies among people with disabilities in two countries in the future. 

Although smart devices helped people with disabilities throughout IADL, there are also concerns related to these smart devices. We found that most smart devices our participants used are commercial products from the general public without any accessible modifications. Our participants reported that they highly rely on these smart devices every day, and it would be better if these products can be more accessible when completing an activity that contains multiple actions, instead of only a single action. For example, P7 complained about the complex actions of cooking rice every day:

\begin{quote}
    ``...I am happy that my rice cooker allows me to start cooking and receive notifications through my phone. However, the hard task for me is to find my uncooked rice, measure the amount, and add the appropriate amount of water before cooking. If all of these procedures could be automatic, it would save me lots of time and effort...''
\end{quote}

In our study, we found that most participants rely on accessibility features, such as screen reader or voice-to-text software, to use mainstream technologies. However, most participants mentioned that the slow update on accessibility features affected the use of mainstream technologies. Specifically, participants with visual impairments complained that the third-party applications' fast updates are not always compatible with their screen readers and can render their screen readers useless. This problem becomes more critical as some people depend on these mainstream technologies for most of the daily activities. Currently, although people with visual impairment could use certain apps to communicate with their friends and shop online, many apps (e.g., games, augmented reality, and virtual reality) are still not accessible to certain groups of people. Future work should have a mechanism to detect the accessibility of software and libraries, especially when the software and libraries are updated.

\section{LIMITATION}

In this work, we studied practices and challenges surrounding ATs use as well as the perceptions and misperceptions of ATs from the perspectives of people with disabilities in China. Most of our participants were from the same province in China. Thus, our participant sample may not generalize to the 83 million people with disabilities across all of China. However, we do think that our work provides an understanding of these matters from a specific context in China, while existing research on these issues have been conducted primarily in North America and Europe only.
An additional limitation is that it is also essential to understand the perceptions and misperceptions of ATs from people who do not have disabilities. 
These two complementary perspectives could provide a more holistic picture to understand the issues that prevent ATs from being socially acceptable. Toward this end, it is worth exploring the perspectives of people without disabilities about ATs and contrast the findings with that of our study as well as prior research to better understand how to make ATs more socially acceptable and how to make people with disabilities feel more inclusive.  

We intentionally interviewed people with a wide range of disabilities to cover a broader range of practices and challenges that they have encountered when using ATs. Despite the broader coverage of the types of disabilities, we still have not yet covered all possible disabilities and different severity levels of the disabilities, such as people with cognitive disabilities and the different severity levels of physical disabilities. The particular type of disability and its severity level could shape the way in which people with such a disability use and perceive their ATs. As a result, future work should extend our current study to cover a broader range of disabilities and levels of severity to examine whether and to what extent the findings still hold or need to be extended. Furthermore, due to our limitation of recruitment, we were only able to recruit participants who had experiences with ATs. However, some people with disabilities might have never used ATs in the first place, and it is important to explore their reasons in future research.

This study revealed the practices and challenges of ATs use and perceptions and misperceptions surrounding ATs in China to some extent. Although we have contrasted our findings with related work that was conducted in North America or other regions, we have not conducted a thorough comparative study to systematically compare the similarities and differences of these issues between China and other countries. Thus, even though the similarities and differences found through our study shed light on this line of research, they are by no means exhaustive. Future work should conduct comparative studies to more systematically compare the similarities and differences in the use, perception, and misperceptions of ATs across countries and cultures.

\section{CONCLUSION}

We have presented a semi-structured interview study conducted in China with 26 people who have various disabilities to understand the practices and challenges surrounding the use of ATs and their perceptions and expectations of ATs. We found that participants with disabilities choose AT devices and ``DIY'' their ATs to signal to others their remaining abilities while also considering functional and financial constraints. Our study further identified the challenges that participants encountered with using ATs in social interactions. It is not uncommon for the general public to misuse infrastructures that are designed to help people with disabilities, use traditional stigmatization terms, offer unwanted help, or even pose safety threats. For the few who worked, their working environments also unintentionally posed challenges. Lastly, we also reported the misperceptions that people with disabilities felt others hold about their use of ATs. Specifically, our findings confirmed that previously found misperceptions in a North American context \cite{shinohara2011shadow} also pervade in China today: ATs could functionally eliminate a disability. Moreover, we found additional misperceptions: ATs should overcome inaccessible social infrastructures; ATs are symbols of permanent disabilities; and ATs with more functionalities are better.

Based on these findings, we recommend designers and researchers to take the following aspects into consideration when designing ATs: consider the advantages of both manual and electronic designs; consider both multi-functional and single-functional designs; understand users' personalization needs and use emerging prototyping technologies (e.g., 3D printing, emerging fabrication) to better integrate aesthetics and customization into AT design; consider in-situ use of ATs and make ATs (both hardware and software) easy to update. Additionally, we found the following two issues that the general public needs to address: avoid misuse of the infrastructures designed for people with disabilities; and learn to offer help to people with disabilities appropriately by considering the ``social'' aspects of ATs. Furthermore, we offer the following design considerations to reduce misperceptions that people with disabilities hold about themselves: build online and offline platforms dedicated to people with disabilities, such as discussion forums and live-streaming platforms, to engage them to communicate and share knowledge and skills that they could learn so that others with similar disabilities could be aware of and encouraged by what they could actually learn and do.

\begin{acks}
We would like to thank Hebei Disabled Persons’ Federation and Shijiazhuang Shi Disabled Persons’ Federation, in particular the following people, for their help in recruitment: Jian Gao, Hongtao Zhen, Yunfeng Shi, Jie Zheng, Yang Xue and Anshi Yu.
This work was supported in part by the Natural Sciences and Engineering Research Council of Canada (RGPIN-2016-06326). 
\end{acks}

\bibliographystyle{ACM-Reference-Format}
\bibliography{main}

\end{document}